\def\>{\rangle}
\def\<{\langle}
\def\br{{\bf r}}
\def\S{{\bf S}}
\def\R{{\bf R}}
\def\Rp{R^{\prime}}
\def\aza{C$_{59}$N}
\def\bucky{C$_{60}$}
\def\cal{}

\documentclass{elsart_hacked}

\usepackage{graphics}
\usepackage{amssymb}
\usepackage{amsmath}

\begin{document}

\begin{frontmatter}



\title{Accurate hyperfine couplings for \aza}


\author[csg]{G\'abor Cs\'anyi\corauthref{cor1}}
\ead{gc121@phy.cam.ac.uk}
\author[taa]{T. A. Arias}
\corauth[cor1]{Corresponding author. Fax: +44 1223 337356}
\address[csg]{TCM Group, Cavendish Laboratory, Univeristy of Cambridge, Madingley Road, CB3 0HE, United Kingdom}
\address[taa]{Laboratory for Atomic and Solid State Physics, Cornell University, Ithaca, NY 14853, USA}

\begin{abstract}
% Text of abstract
We identify the shortcomings of existing {\it ab initio} quantum chemistry
calculations for the hyperfine couplings in the recently characterized
azafullerene, \aza.  Standard gaussian basis sets in the context of
all--electron calculations are insufficient to resolve the spin
density near the cores of the atoms.  Using the Projector Augmented
Wave method implemented on top of a standard pseudopotential plane--wave
density--functional framework, we compute significantly more accurate
values for the Fermi contact interaction.
\end{abstract}

\end{frontmatter}

% main text
%\section{}
%\label{}
%
% ESR
%

Electron spin resonance (ESR) is one of the key tools for structural
studies of defects and radicals.  The hyperfine splitting of the
resonance line by magnetic nuclei is used to identify particular peaks
with various sites in the system.  To make such identifications, a
theoretical prediction of the hyperfine splitting is necessary.  The
performance of such predictions has been historically relatively difficult,
because the values of the spin density, which is the required observable,
are very small and thus hard to resolve.  This problem is much more
pronounced in the cases where the electronic states associated with the
unpaired electron are $p$--like, which further contributes to the
reduction of the absolute spin density values near the ionic cores.

In this paper, we investigate the cause of the especially poor performance of
the electronic structure calculation for the case of the azafullerene,
\aza, and provide a method that predicts significantly more accurate
values for the hyperfine splittings.

%
% background for aza
%

The study of doped fullerenes and nanotubes is a very active area, because
dopants can make fullerenes chemically and electronically active.
The potential applications range from novel semiconductors to nanomechanical
devices and even biologically active agents. A general review
of heterofullerenes is in \cite{Hummelen}.

\begin{figure}
\begin{center}
\scalebox{0.5}{\includegraphics{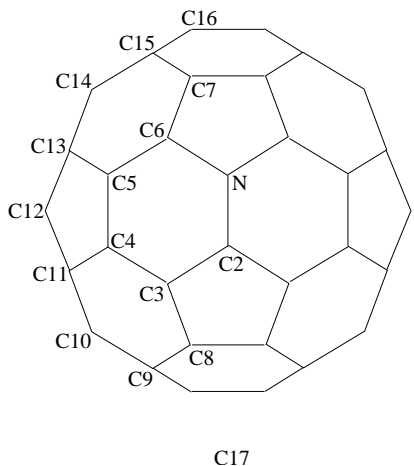}}
\end{center}
\caption{Structure of azafullerene \aza}
\label{fig:c59nstruct}
\end{figure}

In particular, azafullerene, the nitrogen substituted version of
\bucky~has received considerable attention.  The structural and
electronic properties of the
\aza~monomer have been studied before\cite{Andreoni}, and it was found
that there is significant delocalization of the unpaired electron over
the molecular cage.  The degree of delocalization over the \aza~cage
has a significant bearing on the electrical properties of the
material.  However, until recently, \aza~was only available in
dimerized, inactive form.  It is now possible to make solid solutions
of the azafullerene monomer in conventional \bucky~in macroscopic
quantities\cite{andras}, and thus direct comparisons with the
experiments can be made.  The structure of the molecule is shown in
Figure~\ref{fig:c59nstruct}.  The molecule has a mirror plane,
containing the atoms N and C2.

A calculation of the electronic structure of \aza~was published in
\cite{andras} along with the experimental results for the hyperfine
splittings.  The calculated relative values of the hyperfine coupling
were showing the same trends as the experimental data, but the
absolute values were off by more than a factor of two.

%
% explain hyperfine
%

At room temperature, due to the rapid reorientation of the molecules
in the crystal, only the isotropic hyperfine term is observable.  Also
called the Fermi contact interaction, it can
be expressed for a particular nucleus by the effective Hamiltonian
\begin{equation}
H_{\rm hf} = A_{\rm hf} \S_I\cdot\S_e \delta(\br-\R_I),
\label{hfdef}
\end{equation}
where the subscript $I$ refers to the nucleus, and $A_{\rm hf}$ is the
isotropic hyperfine coupling constant and $\S$ is the spin operator.
Because of the Dirac delta function, the hyperfine splitting then only
depends then on the electronic spin density at the position of the
nucleus and the nuclear and electronic magnetic moments,
\[
A_{\rm hf} = \frac{2\mu_B}3 \gamma_e\gamma_I \tilde n(\R_I),
\]
where $\mu_B$ is the Bohr magneton, $\gamma_e$ and $\gamma_I$ are the
electronic and nuclear magnetic moments, respectively, and $\tilde
n(R)$ is the electronic spin density at the nucleus.  The isotropic
hyperfine coupling is thus very sensitive to the values of the
wavefunctions near the atomic cores, which in this case, similarly to
other conjugated systems, can be very small.  There are two sources of
error: the inaccuracy of the density functional used to describe
electronic correlations and the incompleteness of the basis set.  Due
to the aforementioned small numerical value of the spin density at the
nuclear positions, this latter error can be particularly severe.  A
distinct disadvantage of traditional quantum chemistry methods, such
as the one used in \cite{andras} is the uncontrolled nature of the
standard basis sets with which the wavefunctions and the density is
expanded.

Building on the formal work of Hohenberg, Kohn and Sham \cite{HK,KS},
and the later algorithmic work of Car and Parrinello \cite{CPMD} and
others \cite{DFTPW-RMP}, it is now practical to study large systems
with near quantum chemical accuracy using a plane wave basis set.
Below, we use the traditional density--functional plane--wave
pseudopotential method and the local spin density approximation (LSDA)
to relax the molecular geometry, then apply the Projector Augmented
Wave (PAW) method\cite{PAW}, to calculate all--electron spin densities
at the nuclear sites, from which the hyperfine couplings may be
computed.  A key aspect of this approach is that PAW is used as a {\em
post processing} step, thus already existing standard plane--wave
codes can be used to carry out the computationally intensive
geometrical optimization steps.  After that, the hyperfine constants
can be simply and efficiently computed, even using standard numerical
packages.  The {\em ab initio} calculations were carried out with the
DFT++ package \cite{DFT++}, using periodic supercell technique in a
$24$ Bohr cubic cell; the plane--wave cutoff was $20$ Hartrees.  The
convergence of the results with respect to the plane--wave cutoff were
checked by repeating the calculations at $35$ Hartrees; the hyperfine
couplings only changed by less than $10\%$.  We used optimized norm
conserving pseudopotentials developed by Rappe\cite{rappe}.

The PAW method was introduced by Bl\"ochl in 1994. It allows the
reconstruction of all--electron properties from pseudopotential
calculations.  Originally, PAW was presented as a fully
self--consistent all electron method, where the reconstructed
wavefunctions are used to compute the total energy.  Here, we just use
the method of reconstruction.  Accordingly, we define angular momentum
projector functions $|p\>$ for every atomic species, which have the
property
\[
\<p_i|\phi^{\rm ps}_j\> = \delta_{ij},
\]
where $\phi^{\rm ps}_j(\br)$ are atomic pseudo wavefunctions and $i$,
$j$ are composite indices of the usual angular momentum channels 
$[nlm]$.  The projectors $|p\>$ are localized near the nucleus, and they
vanish outside the cutoff radius of the pseudopotential.  The
reconstruction formula is
\begin{equation}
|\Psi^{\rm rec}\> = |\Psi^{\rm ps}\> + \sum_R\sum_i 
\left(|\phi^{\rm ae}_{R,i}\>-|\phi_{R,i}^{\rm ps}\>\right)\<p_{R,i}|\Psi^{\rm ps}\>,
\label{psirec}
\end{equation}
where $|\Psi^{\rm ps}\>$ is an extended pseudo wavefunction,
$\phi^{\rm ae}$ are the orbitals from an all--electron atomic
calculation, and $R$ runs through the position of the nuclei.  Note
that because the atomic pseudo wavefunctions are identical with the
all--electron atomic wavefunctions outside the cutoff radius, the
reconstructed wavefunction $|\Psi^{\rm rec}\>$ is identical to the
pseudo wavefunction $|\Psi^{\rm ps}\>$ in the inter--atomic region.
The reconstruction is exact within the frozen core approximation if we
take a complete set of angular momentum projectors.  In practice, it
is often enough to take just one projector for every
pair $l$ and $m$.

For our purposes here, we also need the notion of a {\em pseudo
operator}, which arises naturally within the PAW formalism.  Using the
definition (\ref{psirec}), the matrix elements of an operator $\cal O$
between reconstructed wavefunctions are given by
\begin{equation}
\<\Psi_1^{\rm rec}|{\cal O}|\Psi_2^{\rm rec}\> =  
\<\Psi_1^{\rm ps}|{\cal O}|\Psi_2^{\rm ps}\> +
\sum_{R\Rp,ij}\<\Psi_1^{\rm ps}|p_{\Rp,j}\>
\left(\<\phi^{\rm ae}_{\Rp,j}|{\cal O}|\phi^{\rm ae}_{R,i}\>-
\<\phi_{\Rp,j}^{\rm ps}|{\cal O}|\phi_{R,i}^{\rm ps}\>\right)
\<p_{R,i}|\Psi_2^{\rm ps}\>.\label{operatoreq}
\end{equation}
Note that for local and semi--local operators, only on--site terms
contribute, where $R=\Rp$.  Thus we can define a pseudo operator $\cal
O^{\rm ps}$,
\[
{\cal O^{\rm ps}} = {\cal O} +\sum_{R,ij}|p_{R,j}\>\left(\<\phi^{\rm ae}_{R,j}|{\cal
O}|\phi^{\rm ae}_{R,i}\>-\<\phi_{R,j}^{\rm ps}|{\cal O}|\phi_{R,i}^{\rm ps}\>\right)\<p_{R,i}|.
\]
This pseudo operator, when acting on pseudo wavefunctions, by
construction will give the same matrix elements as the corresponding
all--electron operator acting on all--electron wavefunctions.

To extract the hyperfine splitting from an {\em ab initio}
calculation, we need the all--electron spin density at the nuclear
positions,
\[
\tilde n(R) = n_\uparrow(R)-n_\downarrow(R).
\]
The simplicity of the delta function operator in (\ref{hfdef}) made this
one of the first applications of the above reconstruction ideas.  If
the angular momentum expansion were complete, the first and third
terms of equation \ref{operatoreq} would cancel exactly, and only the
term involving the all--electron atomic states would need to be
calculated.  Also, only the $s$ states have a non--vanishing
density at the nucleus of an atom, thus (\ref{operatoreq}) reduces to
\begin{equation}
n^{\rm rec}(R) = \sum_i n^{\rm ps}(R) w_i(R)
\label{fullscaling}
\end{equation}
where $i$ is the index of the projectors for the $s$ channels, and the 
weighting factors $w_i$ are
\[
w_i(R) = \frac{n^{\rm ae}_{i,{\rm at}}(0)}{n^{\rm ps}_{i,{\rm at}}(0)},
\]
where the subscript ``at'' refers to densities in isolated atoms.
Bl\"ochl calculated hyperfine constants for a variety doped
semiconductor systems \cite{BlochlHyperfine} using
equation~\ref{fullscaling} and obtained satisfactory results by just
taking a single projector,
\begin{equation}
n^{\rm rec}(R) = n^{\rm ps}(R) w_0(R).
\label{scaling}
\end{equation}
Equation~\ref{scaling} amounts to a simple rescaling of the pseudo
spin density by a constant factor.

The problem with the \aza~molecule, as mentioned above, is that the
unpaired electron is in an orbital which has almost $p$-like symmetry
near the nuclei.  The unpaired spin density at the nucleus is
therefore very small and its value becomes very sensitive to how well
the charge density is resolved near the nucleus.  We find that a more
robust approach is to retain the projector form from equation
\ref{operatoreq} and use the formula
\begin{equation}
n^{\rm rec}(R) = n^{\rm ae}_{\rm at}(R) |\<\Psi^{\rm ps}|p_{R,s}\>|^2.
\label{PAWoperatorform}
\end{equation}

Apart from the issue of resolution near the ionic core, the truncation
of the angular momentum expansion in (\ref{operatoreq}) implies that
the projector form (\ref{PAWoperatorform}) is expected to
underestimate the spin density, while the simple scaling method
(\ref{scaling}) overestimates it by assuming that each $s$ channel has
the same scaling factor $w_i \equiv w_0$ for a given nucleus.  Note
that this error is in addition to those resulting from an incomplete
spatial resolution of the wavefunction near the core.

\begin{figure}
\bigskip
\begin{center}
\scalebox{0.7}{\includegraphics{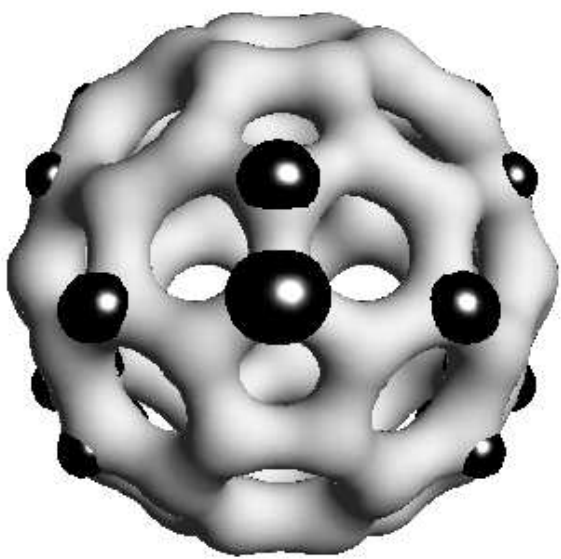}}
\end{center}
\caption{Contour of the charge density (gray) and the spin density
(black). of the \aza~molecule.}
\label{fig:c59n_dens}
\end{figure}

\begin{figure}
\bigskip
\begin{center}
\scalebox{0.5}{\includegraphics{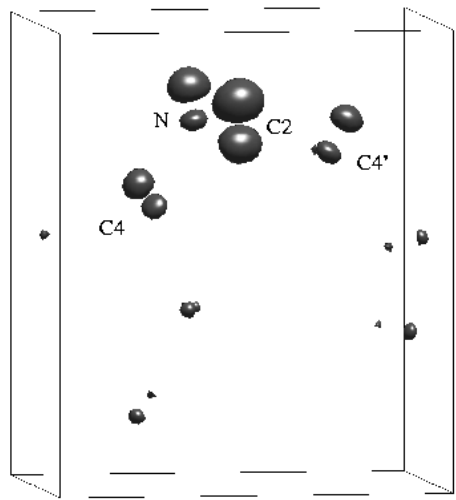}}
\end{center}
\caption{Contour of the spin density in the \aza~molecule.}
\label{fig:c59n_spdens}
\end{figure}

The ionic positions in the \aza~molecule were relaxed until all forces
are less than $0.03~$eV/\AA.  Figure \ref{fig:c59n_dens} shows the
contours of the charge density in gray, and the spin density in black;
Figure \ref{fig:c59n_spdens} shows just the spin density.  The
significant delocalization of the unpaired electron is clearly
visible, as well as the oscillations of the spin density as the
distance increases away from the nitrogen atom.  Table
\ref{table:c59n_hf} shows our computed hyperfine coupling constants
for selected atoms and the results using the {\sc gaussian} all--electron
program\cite{gaussian98}  which were published along with the experiment in
\cite{andras}.  It is clear that the full projection formula
(\ref{PAWoperatorform}) is much more accurate; the error is at most
$20\%$ compared to experiment.  It is interesting that the values
obtained by simply rescaling the pseudo density are relatively close
in agreement with the values from the {\sc gaussian} calculation.
Both traditional approaches overestimate the hyperfine coupling by
over a factor of two.  Note that the experiment only measures the
absolute value of the hyperfine coupling and that the identification
of the experimental values with particular sites is actually inferred
from the calculated values.  Recently, defects in silica were
investigated by Bl\"ochl\cite{BlochlSilica}, where PAW was used as
an all electron method.  There, the conclusion was similar to ours:
much accuracy can be gained by using the full projection method.

In conclusion, we have computed isotropic hyperfine coupling constants
for the azafullerene \aza~and obtained a much better agreement with
recently published experimental data than other calculations.  We
showed that the hyperfine parameters are very sensitive observables in
this system because the unpaired electron is more or less in a
$p$--like state with relatively small spin density at the nuclear
sites.  Thus, a full projection--based formula of the PAW method was
needed to obtain reasonable values for the hyperfine coupling, as
opposed to the currently accepted method of simply rescaling of the
pseudo density.  This underscores the significance of basis set
convergence and related numerical issues regarding the traditional
quantum chemistry basis sets.

\begin{table}
\begin{center}
\bigskip
\scalebox{0.5}{\includegraphics{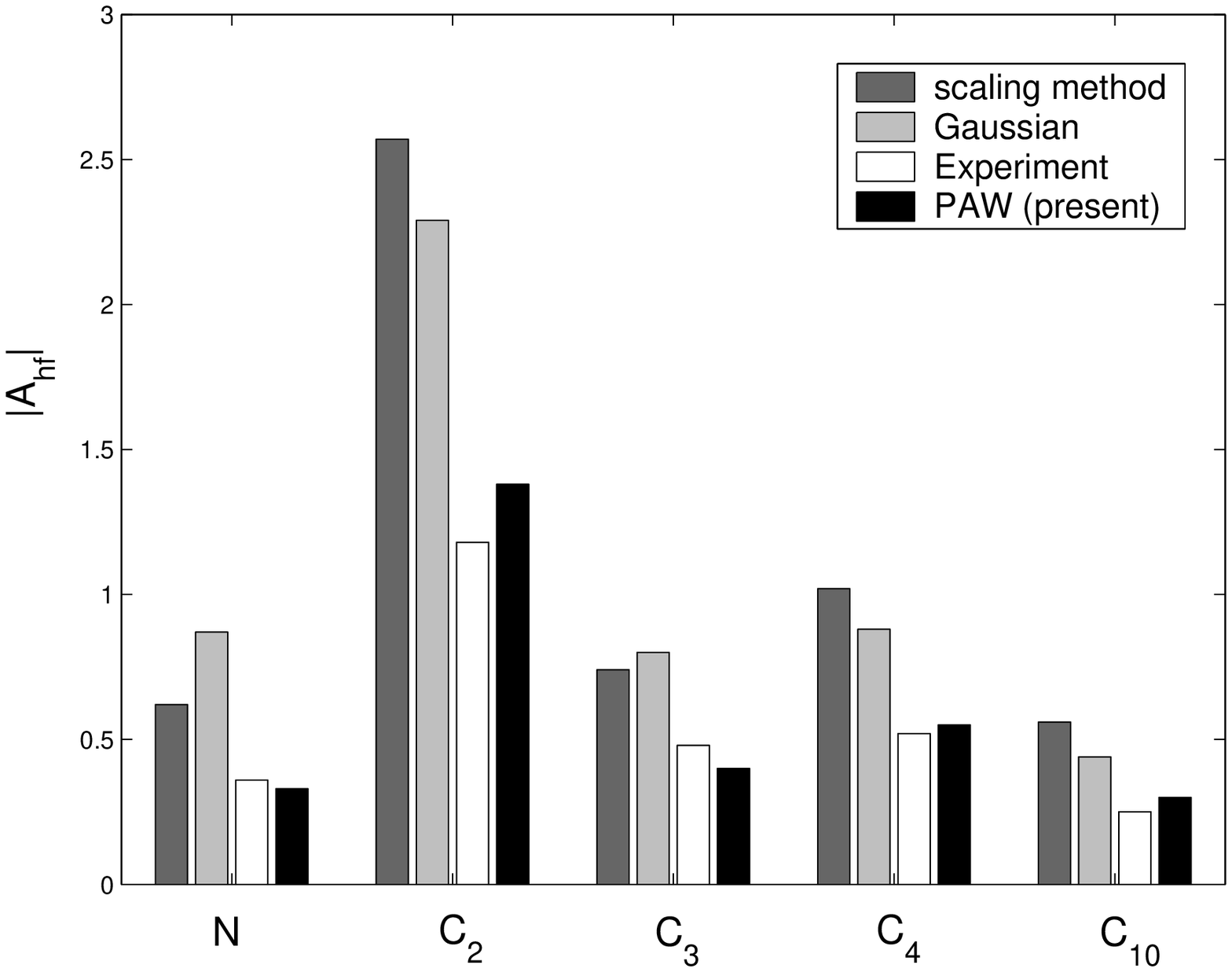}}
\bigskip
\vglue 0.1in
\begin{tabular}{lcccc}
Atom	&Expt.	&PAW projector	&$n^{\rm ps}(R)$ scaling	&{\sc gaussian}\\
	&	&method		&method				&B3LYP\\\hline
N   	&0.36	&  0.33		&0.62				&0.87\\
C2  	&1.18	&  1.38		&2.57				&2.29\\
C3  	&0.48	& -0.40		&-0.74				&-0.80\\
C4  	&0.52	&  0.55		&1.02				&0.88\\
C10 	&0.25	&  0.30		&0.56				&0.44\\
\end{tabular}
\end{center}
\caption{Hyperfine coupling constants in mT.  The present work is
in the second and third column.  The ``PAW projector method'' uses
equation \ref{PAWoperatorform}, the ``scaling method'' uses equation
\ref{scaling}.  Note that the experiment only measures the absolute
value, so this is what is presented in the bar graph.  The {\sc gaussian}
and experimental values are from \cite{andras}.}
\label{table:c59n_hf}
\end{table}

% \appendix

% \section{}
% \label{}

%\begin{thebibliography}{00}


% notes:

% \bibitem{label2}
% If there is a note, it should come last:
% \bibitem{label3} \note
% \end{subbibitems}

%\bibitem{}

%\end{thebibliography}

\bibliographystyle{C59N_els_v2}
\bibliography{C59N_els_v2}% Produces the bibliography via BibTeX.

\end{document}